# The simultaneous measurement of energy and linear polarization of the scattered radiation in resonant inelastic soft x-ray scattering

L. Braicovich,[1,a)] M. Minola,[1,b)] G. Dellea,[1] M. Le Tacon,[2] M. Moretti Sala,[3] C. Morawe,[3] J-Ch. Peffen,[3] R. Supruangnet,[4] F. Yakhou,[3] G. Ghiringhelli,[1] and N. B. Brookes[3]

[1]*CNR-SPIN and Dipartimento di Fisica, Politecnico di Milano, piazza Leonardo Da Vinci 32, Milano, I-20133, Italy*

[2]*Max-Planck-Institut für Festkörperforschung, Heisenbergstrasse 1, D-70569 Stuttgart, Germany*

[3]*European Synchrotron Radiation Facility, 71 Avenue des Martyrs, Grenoble F-38043, France*

[4]*Synchrotron Light Research Institute, Nakhon Ratchasima, Thailand*

Resonant Inelastic X-ray Scattering (RIXS) in the soft x-ray range is an element-specific energy-loss spectroscopy used to probe the electronic and magnetic excitations in strongly correlated solids. In the recent years, RIXS has been progressing very quickly in terms of energy resolution and understanding of the experimental results, but the interpretation of spectra could further improve, sometimes decisively, from a full knowledge of the polarization of incident and scattered photons. Here we present the first implementation, in a high resolution RIXS spectrometer used to analyze the scattered radiation, of a device allowing the measurement of the degree of linear polarization. The system, based on a graded $W/B_4C$ multilayer mirror installed in proximity of the CCD detector, has been installed on the AXES spectrometer at the ESRF; it has been fully characterized and it has been used for a demonstration experiment at the Cu $L_3$ edge on a high-Tc superconducting cuprate. The loss in efficiency suffered by the spectrometer equipped with this test facility was a factor 17.5. We propose also a more advanced version, suitable for a routine use on the next generation of RIXS spectrometers and with an overall efficiency up to 10%.

---

[a)] Author to whom correspondence should be addressed. Electronic mail: lucio.braicovich@polimi.it.
[b)] Now at Max-Planck-Institut für Festkörperforschung, Heisenbergstrasse 1, D-70569 Stuttgart, Germany.

## I. INTRODUCTION

Advances in research using inelastic x-ray scattering are tightly interconnected with the progress in the instrumentation. Together with detection efficiency, high energy and momentum resolution is the main requirement for measuring the momentum dependent excitations of interest by x-ray energy loss spectroscopy. In resonant inelastic x-ray scattering (RIXS)[1,2] the photon energy corresponding to a core level makes RIXS site and chemically specific. In this context a crucial issue is the control of the polarization of x-rays with a further increase of selectivity. As far as the incident beam is concerned, the problem has been solved by the progress in the insertion devices installed in third generation synchrotron radiation sources. The situation is instead much more delicate in the case of the scattered radiation since it is necessary to make the polarization measurement compatible with the spectral analysis without loss in the energy resolution. This has been done in the hard x-ray RIXS[3,4,5] and magnetic scattering[6,7,8] by exploiting the high reflectivity of Bragg optics based on crystals. On the other hand, in the soft x-ray range, covering the very interesting $L$-edges of $3d$ transition metals, the situation is made more difficult by the low efficiency of optical elements. Indeed the measurement of linear polarization of scattered photons has never been done in a soft-RIXS experiment, whereas it has been realized in the recent years in instrumentation with limited or no energy analysis capabilities[9,10,11,12,13]. The only case of combined polarization and energy analysis was realized by Duda *et al.*[14] who exploited the circular dichroism of the $L_3$ resonant reflectivity from a magnetic Fe mirror to obtain sensitivity to circular polarization. The limiting factors of this pioneering contribution are the rapid energy dependence of the response function and the complete lack of tunability, since the polarization detection is obtained by involving a resonant core-level process. On the contrary, the measurement of linear polarization is directly achievable without making use of resonant processes by exploiting the polarization dependence of reflectivity. Therefore the utilization of a multilayer (ML) mirror is ideal for the soft x-ray range, as shown hereafter.

Here we aim at presenting this problem in a rather comprehensive way by providing the information needed to implement polarization analysis in almost any soft x-ray spectrometer. This presentation is based on two topics logically connected. i) A pilot experiment done with the AXES spectrometer[15,16] at the beamline ID08 of the ESRF, which has led to unique and important results. The equipment is finalized to the Cu-$L_3$ edge (930 eV) but is sufficiently tunable to cover Cu in different compounds and different ionization states. A brief, non-technical account of this project has already been given in ref. 17. ii) The conceptual design of an innovative polarimeter designed for the ultrahigh-resolution ERIXS spectrometer to be operated at the new beamline ID32 of the ESRF from 2015; this instrument takes advantage of the pilot experiment.

The choice of the Cu $L_3$ edge is due to the importance of high-$T_c$ cuprate superconductors. These systems have important magnetic excitations at low energy and the problem of their detailed identification is crucial and RIXS has recently demonstrated to be a very valuable alternative to inelastic neutron scattering[18] in this field[19,20]. These specific experiments will particularly benefit from the measurement of the linear polarization, since the magnetic excitations are accompanied by 90 deg rotation of the polarization vector of the scattered photons with respect to the incident ones[21].

## II. BASIC CONCEPTS

### A. Obtaining the sensitivity to polarization

Due to the low counting rate, which is often the main limiting-factor, in soft-RIXS spectra are acquired with a parallel detection of all energies within a several-eV window. Thus in all instruments a common fact is the angular dispersion (in the vertical direction) of the radiation operated by a diffraction grating and the detection of the photons by a position sensitive detector (usually a micro-channel plate or a CCD). Irrespective of the number of optical elements, the optical axis, in the absence of horizontal focusing, projects onto a straight line in a top view as in Fig. 1(a) and the beam diverges horizontally from the sample to the detector. The addition of a focusing mirror can reduce the horizontal divergence as shown in Fig. 1(b): the most interesting case of horizontal focus is the collimation with a parabolic mirror (focus at infinity). In all these cases the idea is to make the system sensitive to linear polarization by inserting a final optical element (in green in the cartoon of Fig. 2, referring as an example to a system with only one optical element). This additional element ideally deflects the beam without affecting it in the (vertical) dispersion direction. It is a multilayer mirror working in Bragg conditions at the desired photon energy with intermediate incidence angle, so that its reflectivity is larger with σ' polarization (electric vector **E** in the plane of the multilayer) than with π' polarization (**E** in the scattering plane). Note that, for consistency along the whole article, we will refer to the polarization components perpendicular and parallel to the scattering (reflection) plane as σ and π (σ' and π') respectively before (after) the sample. We anticipate here that, although the effect is strongest at 45 deg (the Brewster angle for x-rays), and lowest at normal and very grazing incidence, already around 20 deg from the mirror surface the polarization dependence of reflectivity is strong enough for our purposes.

**B. Extracting fully polarized spectra**

Without loss of generality, let us refer to an instrument as in Fig. 2 where the ML can be inserted in the beam just before the detector and can be rotated around the *aa* axis contained in the ML surface. For the polarization analysis two spectra are needed, measured with identical energy resolution and experimental conditions: spectrum $I_D$ is obtained by the direct detection, spectrum $I_M$ is recorded after the reflection of the photons by the ML mirror. By conveniently combining spectra $I_D$ and $I_M$, thanks to the accurate knowledge of the reflectivity for the two polarizations σ' and π', it is possible to recover the two linear components of the polarization (in the assumption of a fully polarized beam). The cases σ' and π' are defined with respect to the ML, i.e. with the electric vector parallel or perpendicular to the ML surface respectively. This reference differs from the sample (or laboratory) reference system due to the inclination of the output arm. Since this angle is small (less than 8 deg) this rotation has a negligible effect when compared to statistical error bars and will be neglected. Thus we will identify the σ' polarization with **E** vertical and π' with **E** horizontal in the laboratory frame. As already mentioned, the prime (') index denotes the polarization directions of the photons scattered from the sample, whereas σ and π refer to the radiation impinging onto the sample.

The system as a whole can be regarded as a filter with a small efficiency (the average reflectivity of the ML mirror is typically around 10%) and with preferential transmission of the σ' component with respect to π'. Thus the $I_M$ spectra differ from $I_D$ spectra for two reasons: the overall intensity is decreased by the reflectivity of the ML mirror, and the relative weight of the σ' and π' components has been modified in the reflection. Formally, let us define the reflectivity of the multilayer for the two polarizations as:

$$r_{\pi'}(h\nu) = [1 - \delta(h\nu)]r(h\nu) \qquad (1)$$

$$r_{\sigma'}(h\nu) = [1 + \delta(h\nu)]r(h\nu) \qquad (2)$$

where $r(hv)$ is the reflectivity averaged over the two polarizations, and $\delta$ measures the polarimetric effect. Then, following the definitions given above,

$$I_D(hv) = I_{\pi\prime} + I_{\sigma\prime} \tag{3}$$

$$I_M(hv) = F[r_{\pi\prime}I_{\pi\prime} + r_{\sigma\prime}I_{\sigma\prime}] = F\big[[1-\delta]I_{\pi\prime} + [1+\delta]I_{\sigma\prime}\big]r \tag{4}$$

where $F$ accounts for the fraction of the beam intercepted by the ML mirror and for the other losses of average intensity in the real system. We obtain then:

$$I_{\pi\prime}(hv) = \frac{I_D + I_\Delta}{2} \tag{5}$$

$$I_{\sigma\prime}(hv) = \frac{I_D - I_\Delta}{2} \tag{6}$$

with

$$I_\Delta(hv) = I_{\pi\prime} - I_{\sigma\prime} = \frac{1}{\delta}\left(I_D - \frac{I_M}{Fr}\right) \tag{7}$$

The $I_\Delta$ spectrum (for one given incident polarization, $\sigma$ or $\pi$) has a clear meaning. In fact $\frac{I_M}{Fr}$ is the spectrum corrected for the average reflectivity of the multilayer and for intensity losses. In the absence of any polarization effects, $\frac{I_M}{Fr} = I_D$ so that $I_\Delta = 0$. Therefore the polarization effect is represented by the departure from zero of $I_\Delta$, amplified by the factor $1/\delta$ that accounts for the sensitivity to the polarization.

It must be considered, though, that the polarimeter can be used only if one knows, in the actual working conditions, the energy dependence of $Fr$ and $\delta$. In our system, as initial guideline, the functions $r(hv)$ and $\delta(hv)$ were measured with a diffractometer installed on the same beamline. These measurements were done at one point of the ML, without scanning the position, due to technical constraints. We will show in section IV how to measure them in situ, i.e., in realistic conditions. The diffractometer data give the peak of the function $r$ around 0.095 and our geometrical acceptance factor $F$ is 0.68, so that the reduction of the counting rate ($1/Fr$) is expected to be approximately 15.5. This is tolerable in a modern beamline with high luminosity but these measurements remain in any case very delicate.

## III. THE PILOT INSTRUMENT

### A. The spectrometer

The device presented below has been realized for the AXES spectrometer[15]; after several refurbishments[16,22] this instrument, based on a variable line-density spherical grating and a CCD detector, has at present a resolving power $E/\Delta E$ = 4000-4500 (at 930 eV), corresponding to about 3500 when combined with the beamline monochromator. As shown in Fig. 2, the whole spectrum is acquired simultaneously, without scanning parts, simply by accumulating for several minutes the signal forming an intensity image on the CCD, and by integrating along the (horizontal) lines of constant photon energy (slightly curved due to aberrations generated by the horizontally diverging beam). The detector is inclined so that the incidence angle on its surface is small (in this case 16 deg) to decrease the effect of the finite spatial resolution of the CCD. The total length of the system (from sample to detector) is 2.2 m. The beam coming from the beamline is focused on the sample so that the vertical height is less than 10 μm and acts as a virtual input slit.

### B. Optical matching between polarimeter and spectrometer

This is the crucial issue. In fact the above estimate of the counting rate is based on the assumption that each illuminated point of the ML mirror contributes to the spectrum with equal efficiency. This ideal matching is non-trivial since the horizontal acceptance of the spectrometer is much larger than the angular window accepted by a standard, plane ML. With reference to Fig. 3, a ML with fixed period $d_0$ would reflect with good efficiency only from a narrow central area where the incidence angle $\theta_0$ is optimized for the selected energy $E$; due to the divergence of the beam (up to 10 mrad in AXES) in the lateral areas the incidence angles ($\theta_1$ and $\theta_2$ in the cartoon) would be too different from $\theta_0$ and the reflectivity would drop quickly, thereby reducing decisively the average detection efficiency when using the ML mirror. The solution is a ML with spacing *d* variable in the horizontal direction of the mirror so that the same energy is in Bragg condition in each point of the ML along the horizontal fan. Simulations and our preliminary experimental work show that in AXES a graded multilayer recovers a factor of 4-5 in intensity with respect to a constant-period ML. Moreover the simulations have shown that a grading linear with the position is an excellent approximation. In an ideal case without defects of the ML and with perfect alignment, this optical matching ensures that the energy resolution is not affected.

The graded ML has been installed as close as possible to the CCD in order to minimize the effect of possible stray light and to be compatible with the pre-existing apparatus: the distance to the CCD is 25 cm. For space reasons the mirror size had to be limited and only 68% of the horizontal fan could be intercepted. These limitations would not be present in a totally new apparatus and are acceptable in the spirit of a pilot project. During operations what really matters is the control of the increment of the rotation angle $\theta$ of the ML. In our instrument we control increments of $\theta$ down to 1 μrad. A further constraint is the limit to the total deflection ($2\theta$) by the ML. This is around 40 deg, which is smaller than the optimal statistical value of 48 deg given by the simulations. We stress that much larger deflections, including $2\theta = 90$ deg that gives $r_{\pi'} = 0$, are incompatible with the already scarce intensity of RIXS experiments, although they are still barely sufficient for diffractometers[12]. For example, simulations[23] for a W/B$_4$C ML with an excellent 0.25 nm (rms) inter-diffusion at interfaces, give at 930 eV the reflectivity $r_{\sigma'}(\theta,d)$ of the σ' component drops from $r_{\sigma'}(20\ \text{deg}, 1.93\ \text{nm}) = 1.1 \times 10^{-1}$ to $r_{\sigma'}(45\ \text{deg}, 0.95\ \text{nm}) = 7.5 \times 10^{-3}$: an instrument having 100% polarization sensitivity would have a counting rate unacceptably low.

The problem is less compelling at lower energies, where the period of the ML is larger [at 530 eV, $r_{\sigma'}$(20 deg, 3.38 nm) = $7.8 \times 10^{-2}$; $r_{\sigma'}$(45 deg, 1.67 nm) = $3.8 \times 10^{-2}$] but becomes untreatable at higher energies [at 1190 eV, $r_{\sigma'}$(20 deg, 1.54 nm) = $1.0 \times 10^{-1}$; $r_{\sigma'}$(45 deg, 0.74 nm) = $7.5 \times 10^{-4}$]. A further requirement for our ML is a reflectivity peak as broad as possible, and compatible with an average reflectivity not far from 10%. These performances were optimized using 150 periods made of W and $B_4C$. The gradient in the direction of the beam footprint is 2.3% and the central spacing is 1.91 nm. The peak reflectivity, measured at 930 eV with a diffractometer at ID08, is $r_{\sigma'}$ = 0.112, close to the simulations. The mirror was grown at the ESRF multi-layer fabrication facility by RF sputtering[24] and was characterized in the growth laboratory with a conventional x-ray source at 8048 eV; on this basis the extrapolated values at $L_3$ edge are in agreement with the measurements made later at the beam line.

The mechanical setup allows the ML to move along the axis used in the rotation (axis *aa* in Fig. 2). This vertical translation is used to clear the way to the CCD in the measurement without the polarimeter. This movement can be used also to work at other absorption edges provided different ML stripes are deposited on the same mirror.

## IV. THE PILOT EXPERIMENT: TESTS AND RESULTS

### A. Test experiment

In order to reliably use the performances of the polarimeter the actual reflectivity for the two polarizations $r_{\pi'}$ and $r_{\sigma'}$ had to be measured in the working conditions and with low statistical uncertainty. A normal RIXS spectrum could not be used, due to the weakness of the signal. This difficulty was bypassed by the use of a suitable multilayer instead of the sample: in Bragg reflection conditions the intensity is orders of magnitude higher than in RIXS and it can be easily used for the reflectivity calibration. In our case the Bragg angle was 65 deg to the surface and the ML acting as a sample had even to be slightly detuned to avoid overloading or eventually damaging the CCD detector. With that configuration an almost complete diagnostic could be completed in a few hours. This type of preliminary calibration of the polarimeter should be repeated rather often, typically at the beginning of each experiment or every time the geometrical arrangement of the sample is modified, even slightly.

First of all the optical matching is checked by testing the energy resolution. This is done in a very stringent mode since the high intensity allows working at the maximum resolving power. In this case the combined full width at half maximum (FWHM) of an elastically scattered beam is 240 meV at 931 eV, measured without polarimeter (Fig. 4, black squares); the very same line-width is obtained also when the polarimeter is inserted (red circles). In the two configurations the system has a slightly different energy calibration (about 1.5% difference), taken into account in the figure. The results of Fig. 4 (obtained with a total counting time of 3 minutes) demonstrate that the energy resolution is fully preserved by the polarimeter. This procedure gives also the scaling factor $Fr$ at the energy used in the test.

The next test deals with the energy dependent response function of the system at fixed angle of the ML mirror. The test consists of taking a series of monochromatic lines in steps of 0.5 eV, with σ and π incident polarization, without and with the polarimeter, as shown in panels (a) and (b) of Fig. 5, where the raw data are displayed in counts/minute. The effect of the polarimeter is evident in the reduction of the average intensity and in the marked increase of the difference between the σ (solid lines) and the π case (dots). Note also the slight departure from the usual polarization factor is typical in soft x-rays. After taking into account the actual intensity of the incoming beam during the experiment ($I_0$), we could evaluate the intensity loss due to the polarimeter, $1/Fr = 17.5$, very close to the expected value 15.5 mentioned above. This is satisfactory when considering that the ideal case assumes a perfect filling of the CCD. From the data of Fig. 5, the response functions are easily extracted and are plotted in Fig. 6(a) (red and black symbols) together with the corresponding values obtained previously with the diffractometer (red and black thick solid lines). As an indicator of the polarization sensitivity we plot in black the flipping ratio $(1+\delta)/(1-\delta)$: the two sets of results are in good agreement. Different is the situation of the average reflectivity (in red) normalized to the maximum. We see a substantial agreement in a range about 1 eV wide around the "central" energy of 931 eV, while at lower energies the *in situ* results decrease more rapidly. This happens since one loses intensity at lateral sides of the CCD which has the full filling at fixed θ only at the central energy. This analysis demonstrates that it is mandatory to measure *in situ* the real response function to be used in data analysis.

Moreover the rapid decrease of the reflectivity at lower energies makes the use of the equations of section II risky, due to the heavy data corrections to get the function $I_\Delta$ (see eq. 7). For this reason we restrict the use of the polarimeter to a 2.5 eV energy range below the 931 eV, which is acceptable for most of the cuprates.

Up to here we have been working at constant θ but the performances depend both on the energy and on the angle θ and it is illuminating to consider the response in two dimensions, i.e., as function of both θ and energy. A typical rocking curve is given in Fig. 6(b) (in red), compared with the diffractometer measurements: in blue a θ-scan at fixed 2θ, and in black a θ-2θ-scan. The polarimeter rocking curve is similar to this last case because the CCD is very wide so that a large fraction of the intensity arrives on the detector when the ML is rotated. However, the polarimeter curve decays faster than the θ-2θ-scan as it must, since the detector is not moved during the scan.

This explains the difference between the real and the ideal scaling factor and makes the dependence on the angle much slower than in ordinary θ-scan.

The choice of θ is not trivial, as shown by the map of Fig. 6(c), where the measured intensity with σ' polarization is given vs. energy and angle. To our purpose a map based on four rocking curves is sufficient. The maximum intensity is obtained obviously when the polarimeter is in Bragg conditions. The response functions vs. energy are given by the cuts of the map at constant θ. It is apparent that the response function given by the dashed cut is much smoother than the cut through the Bragg peak. Of course, if everything is consistent, the final result after spectra analysis is the same at different cuts but the reliability of the data is different. The dashed cut is appropriate to measurements extending in a wide energy range, whereas the solid cut is more appropriate to the low energy magnetic excitations, appearing so close to the elastic peak to take full advantage from the reflectivity having been optimized at 931.5 eV. This is the case of all results presented here.

Strictly speaking, in the startup of an experiment, a map is not indispensable but it is greatly useful. At the beamline ID08 (at present dismantled), where this work has been done, the whole set of tests including the map could be completed in about 8 hours. We expect that at the new ID32 beam line these preliminary operations to be faster, easier and needed less often.

## B. A typical result on cuprates

We give here an example of the use of the instrument in the study of cuprates. We chose a case in which it is expected from the theory that the traditional spectrum is the superposition of the two outgoing polarization both with sizable weights[21]. With large scattering angles (in our case 130 deg) this happens at grazing incidence (30 deg in our case) with respect to the surface containing the (110) plane; this expectation is more appropriate at low doping so that we chose the underdoped $YBa_2Cu_3O_{6.6}$[18]. superconductor[25]. A scheme of the experimental geometry is shown in Fig. 7(a). In order to accumulate in five hours a reasonable statistics, we relaxed the resolution to about 320 meV FWHM (combined). The spectrum measured without polarimeter and σ incident polarization is shown in Fig 7(b), where the red box is the region where we use the polarimeter to obtain the results of panel 7(c). The decomposition in fully polarized subcomponents is very clear: the σσ' component (blue) contains the elastic peak broadened at the left by a not resolved phonon contribution and a tiny bimagon shoulder around 500 meV, while the crossed component σπ' (red) is the magnon. The error-bars are the statistical values based on errors propagated through the formulas used to extract the polarized component. The number of accumulated counts of the traditional spectrum is much larger so that the error bars are much smaller and are not reported for readability. With the angles given above the transferred moment in the (100) direction is 0.365 reciprocal lattice units. By rotating the sample it is possible to change this momentum component thus mapping the dispersion of the magnon free of contamination coming from components of different symmetry which are filtered out by the polarimeter.

## V. THE HORIZONTALLY COLLIMATED BEAM: A NEW OPPORTUNITY FOR POLARIZATION MEASUREMENTS

The results presented in the previous section demonstrate that full polarization control in soft-RIXS is possible and interesting. But the low count rate and the energy range covered by the polarization analysis were still restricting the applicability of the technique. Moreover the 2.5 eV range, sufficient for most of the cuprates, would be even smaller for lighter $3d$ transition metals, and thus too narrow. With the setup presented above the expansion of the ML band-pass would imply a further, unacceptable reduction of the counting rate, as shown in Fig. 5. In our case the pilot experiment was realized with the AXES spectrometer at the beamline ID08, recently replaced by the ERIXS spectrometer and the new beamline ID32 respectively, both designed for optimized intensity and ultra-high resolving power[26]. Moreover the spectrometer is equipped with a parabolic mirror collecting a 20 mrad horizontal fan and making it collimated. As a consequence the grading in the horizontal direction is no more needed, because with a parallel beam all the rays impinge at the same angle on the ML mirror. Then a constant-period ML, optimized for a generic energy $E_0$, would work as shown in Fig. 8. A slightly different energy $E_1$ (with $E_1 > E_0$) would impinge on the ML at a different vertical position than $E_0$ due to the dispersing properties of the grating. Then if $E_0$ is at the Bragg condition for the ML, $E_1$ is out of tune and is reflected with lower efficiency than $E_0$, and away from the central energy the intensity past the ML would drop as shown in Fig. 8(a). The band-pass can therefore be broadened by tuning the local period of the ML so to follow the energy dispersion on the multilayer itself dictated by the grating (Fig. 8(b)). This solution is made possible by the use of a graded ML, where the period is linearly modulated in the vertical direction: this is a *dispersion compensating multilayer,* with a period gradient along the dispersion direction. This set-up, that allows covering continuously a wide energy range, will be studied experimentally in the next year, with particular attention to the interplay with energy resolution. A crucial component is obviously the ML, which has been already prepared as explained in ref. 27. This innovative layout is the main perspective in the important and delicate coupling of energy and polarization measurement in soft RIXS.

## IV. CONCLUSIONS

The test experiment realized with the polarimeter installed on the AXES spectrometer demonstrates that it is possible to measure the linear polarization together with the spectral distribution of the radiation in soft-RIXS. Together with the good control of the polarization of the radiation incident on the sample, granted by the modern undulators, this achievement opens the opportunity of performing RIXS experiments in a fully polarized mode. The knowledge of the changes in the polarization state of scattering photons can be illuminating on the type of excitations left behind in RIXS, therefore greatly helping in the spectral assignments. Some of the future spectrometers will be compatible with graded-ML polarimeters capable of covering large energy ranges with flat response, and opening the way to a routinely access to the fully polarized experimental set-up in RIXS. This type of advanced polarimeter for RIXS will reduce the measurement efficiency by at least a factor 10 to 20, affordable on the best beam lines dedicated to RIXS.

## ACKNOWLEDGMENTS

We thank B. Keimer and C. Mazzoli for useful discussions. We also acknowledge L. Eybert for engineering and A. Fondacaro technical assistance. The project was partly supported by the PIK project "POLARIX" of the Italian Ministry of Research (MUIR).

## REFERENCES

[1] A. Kotani and S. Shin, Rev. Mod. Phys. **73**, 203 (2001).

[2] L. J. P. Ament, M. van Veenendaal, T. P. Devereaux, J. P. Hill, J. van den Brink, Rev. Mod. Phys. **85**, 705 (2011).

[3] K. Ishii, S. Ishihara, Y. Murakami, K. Ikeuchi, K. Kuzushita, T. Inami, K. Ohwada, M. Yoshida, I. Jarrige, N. Tatami, S. Niioka, D. Bizen, Y. Ando, J. Mizuki, S. Maekawa, and Y. Endoh, Phys. Rev. B **83**, 241101 (2011).

[4] X. Gao, C. Burns, D. Casa, M. Upton, T. Gog, J. Kim, C. Li, Rev. Sci. Instrum. **82**, 113108 (2011)

[5] K. Ishii, I. Jarrige, M. Yoshida, K. Ikeuchi, T. Inami, Y. Murakami, J. Mizuki, J. Electron Spectrosc. Relat. Phenom. **188**, 127 (2013)

[6] V. Scagnoli, C. Mazzoli, C. Detelfs, P. Bernard, A. Fondacaro, L. Paolasini, F. Fabrizi, and F. de Bergevin, J. Synchrotron Radiat. **16**, 778 (2009)

[7] L. Paolasini, C. Detlefs, C. Mazzoli, S. Wilkins, P. P. Deen, A. Bombardi, N. Kernavanois, F. de Bergevin, F. Yakhou, J. P. Valade, I. Breslavetz, A. Fondacaro, G. Pepellin and P. Bernard, J. Synchrotron Rad. **14**, 301(2007)

[8] C. Detlefs, M. Sanchez del Rio, and C. Mazzoli, Eur. Phys. J. Special Topics **208**, 359 (2012)

[9] L. Braicovich, A. Tagliaferri, E. Annese, G. Ghiringhelli, C. Dallera, F. Fracassi, A. Palenzona, N. B. Brookes, Phys. Rev. B **75**, 073104 (2007)

[10] L. Braicovich, G. van der Laan, A. Tagliaferri, E. Annese, G. Ghiringhelli, N. B. Brookes, Phys. Rev. B **75**, 184408 (2007).

[11] F. Schäfers, H.-C. Mertins, A. Gaupp, W. Gudat, M. Mertin, I. Packe, F. Schmolla, S. Di Fonzo, G. Soullié, W. Jark, R. Walker, X. Le Cann, R. Nyholm, and M. Eriksson, Appl. Opt. **38**, 4074 (1999)

[12] U. Staub, V. Scagnoli, Y. Bodenthin, M. Garcia-Fernandez, R. Wetter, A. M. Mulders, H. Grimmer and M. Horisberger, J. Synchrotron Radiat. **15**, 469 (2008)

[13] T. A. W. Beale, T. P. A. Hase, T. Iida, K. Endo, P. Steadman, A. R. Marshall, S. S. Dhesi, G. van der Laan, and P. D. Hatton, Rev. Sci. Instrum. **81**, 073904 (2010)

[14] L.-C. Duda, P. Kuiper, D. C. Mancini, C.-J. Englund, J. Nordgren, Nucl. Instrum. Methods A **376**, 291 (1996).

[15] C. Dallera, E. Puppin, A. Fasana, G. Trezzi, N. Incorvaia, L. Braicovich, N. B. Brookes, J. B. Goedkoop, J. Synchrotron Radiat. **3**, 231 (1996).

[16] M.E. Dinardo, A. Piazzalunga, L. Braicovich, V. Bisogni, C. Dallera, K. Giarda, M. Marcon, A. Tagliaferri, G. Ghiringhelli, Nucl. Instrum. Methods A **570**, 176 (2007).

[17] G. Ghiringhelli, L. Braicovich, J. Electron Spectrosc. Relat. Phenom. **188**, 26 (2013).

[18] J. M. Tranquada, X. Guangyong, I. A. Zaliznyak, J. Magn. Magn. Mater. **350**, 148 (2014).

[19] L. Braicovich, L. J. P. Ament, V. Bisogni, F. Forte, C. Aruta, G. Balestrino, N. B. Brookes, G. M. De Luca, P.G. Medaglia, F. Miletto Granozio, M. Radovic, M. Salluzzo, J. Van den Brink, G. Ghiringhelli, Phys. Rev. Lett. **102**, 167401 (2009).

[20] L. Braicovich, J. van den Brink, V. Bisogni, M. Moretti Sala, L. J. P. Ament, N. B. Brookes, G. M. De Luca, Salluzzo, T. Schmitt, V. N. Strocov, G. Ghiringhelli, Phys. Rev. Lett. **104**, 077002 (2010).

[21] L. J. P. Ament, G. Ghiringhelli, M. Moretti Sala, L. Braicovich, J. Van den Brink, Phys. Rev. Lett. **103**, 117003 (2009).

[22] V. Bisogni, L. Simonelli, L. J. P. Ament, F. Forte, M. Moretti Sala, M. Minola, S. Huotari, J. van den Brink, G. Ghiringhelli, N. B. Brookes, L. Braicovich, Phys. Rev. B **85**, 214527 (2012).

[23] The Center for X-Ray Optics, Lawrence Berkeley National Laboratory, Materials Sciences Division, X-ray Database: http://henke.lbl.gov/optical_constants/multi2.html

[24] Ch. Morawe, Ch. Borel, J-Ch. Peffen, *The new ESRF multilayer deposition facility*, Proc. SPIE 6705-04 (2007).

[25] M. Le Tacon, G. Ghiringhelli, J. Chaloupka, M. Moretti Sala, V. Hinkov, M. W. Haverkort, M. Minola, M. Bakr, K. J. Zhou, S. Blanco-Canosa, C. Monney, Y. T. Song, G. L. Sun, C. T. Lin, G. M. De Luca, M. Salluzzo, G. Khaliullin, T. Schmitt, L. Braicovich, B. Keimer, Nature Phys. **7**, 725 (2011).

[26] http://www.esrf.eu/UsersAndScience/Experiments/ElectStructMagn/ID32

[27] Ch. Morawe, J-Ch. Peffen, R. Supruangnet, L. Braicovich, N. B. Brookes, G. Ghiringhelli, F. Yakhou-Harris, *Graded multilayers for fully polarization resolved Resonant Inelastic X-ray Scattering in the soft x-ray range,* SPIE (to be published).

**FIGURES**

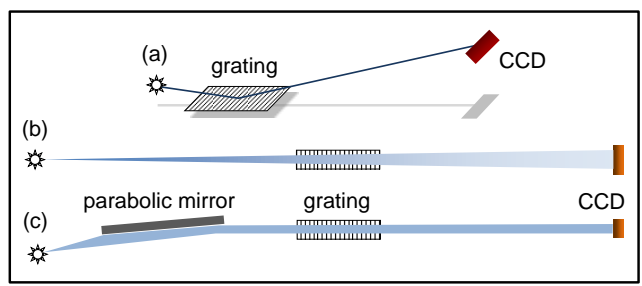

FIGURE 1. Scheme of soft-RIXS grating spectrometer based on a single optical element. Panel (a): perspective view. Panels (b) and (c): top view, in the absence of horizontal focusing (b) and in the case of horizontal collimation obtained with a parabolic mirror (c).

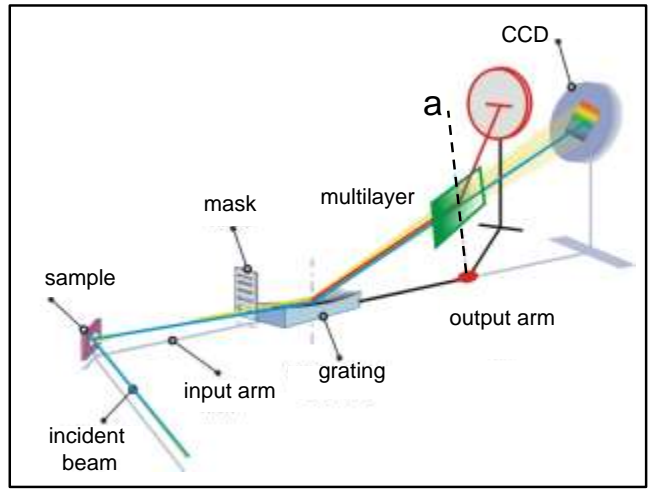

FIGURE 2. Cartoon showing the insertion of a multilayer (in green) along the spectrometer output arm. Both the multilayer and the CCD can be rotated around the a axis, belonging to the multilayer surface and normal to the output arm.

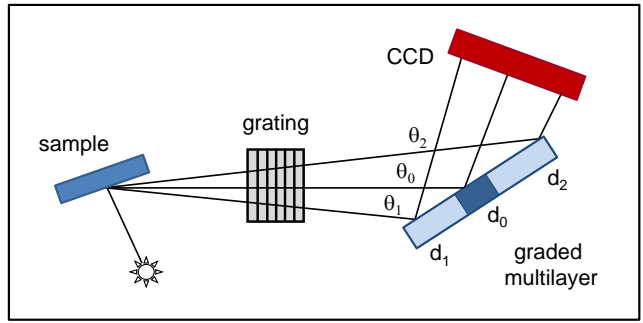

FIGURE 3. Schematic illustration of the need for a graded multilayer with a diverging beam. At constant spacing, if $\theta_0$ is the right angle at energy *E*, only the central region of the multilayer ( dark blue in figure) satisfies the Bragg condition, and angles $\theta_1$ and $\theta_2$ imply low reflectivity. By inserting a ML with a variable spacing (period *d*), the same energy is in Bragg condition in each point of the ML along the horizontal fan and a factor of 4-5 in intensity is recovered.

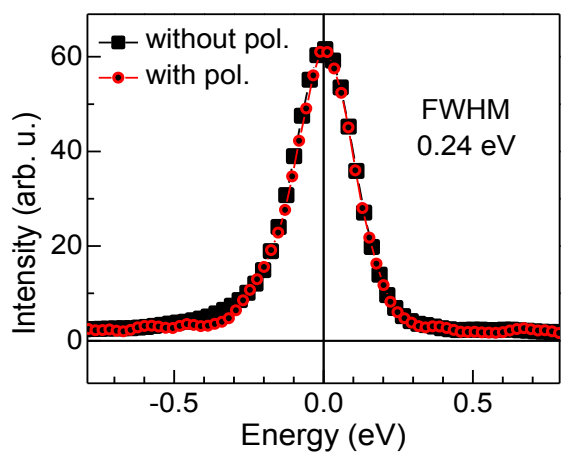

FIGURE 4. Single monochromatic peak as measured using AXES at ID08 of the ESRF, without (black squares) and with (red circles) polarimeter, after normalization to the same maximum intensity. These are raw data where the linewidth includes the contribution from the beamline.

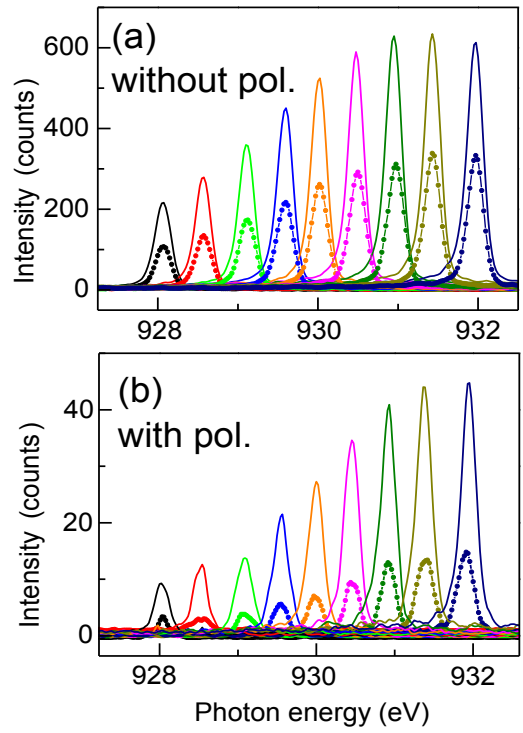

FIGURE 5. Set of measurements (raw data) of monochromatic peaks with incident σ (solid lines) and π (dots) polarization on the sample. Panel (a): data without polarimeter, and panel (b): data with polarimeter.

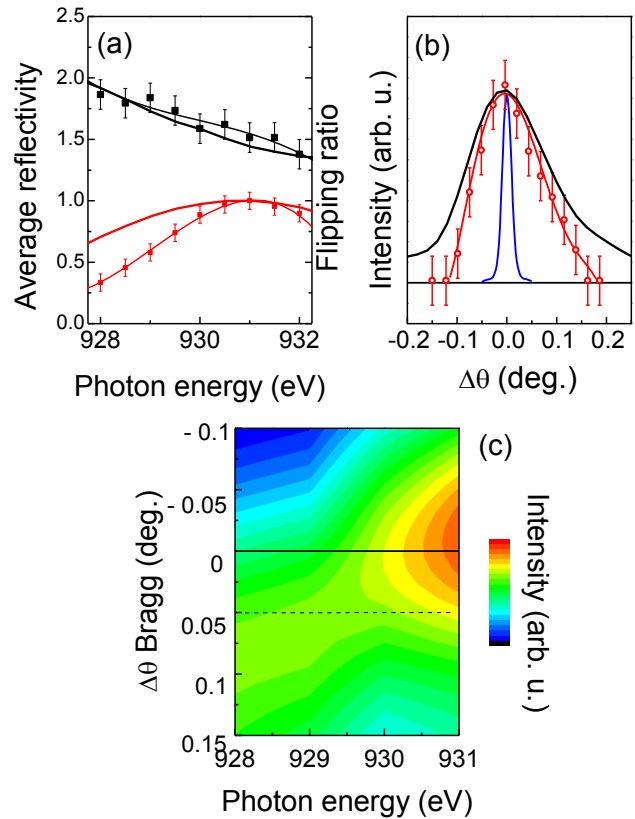

FIGURE 6. Multilayer mirror reflectivity characterization. Panel (a): comparison of the response functions measured *in situ* (dots with thin lines) and *ex situ,* with a diffractometer (thick solid lines). The flipping ratio (1+δ)/(1−δ) is given in black and the average reflectivity normalized to unity in red. Panel (b): rocking curve (θ-scan) of the polarimeter (red open circle with red lines) and the results from the diffractometer (solid lines); the blue line is a θ-scan and the red a θ-2θ−scan. Panel (c): an intensity map vs. incident energy and incremental θ of the multilayer.

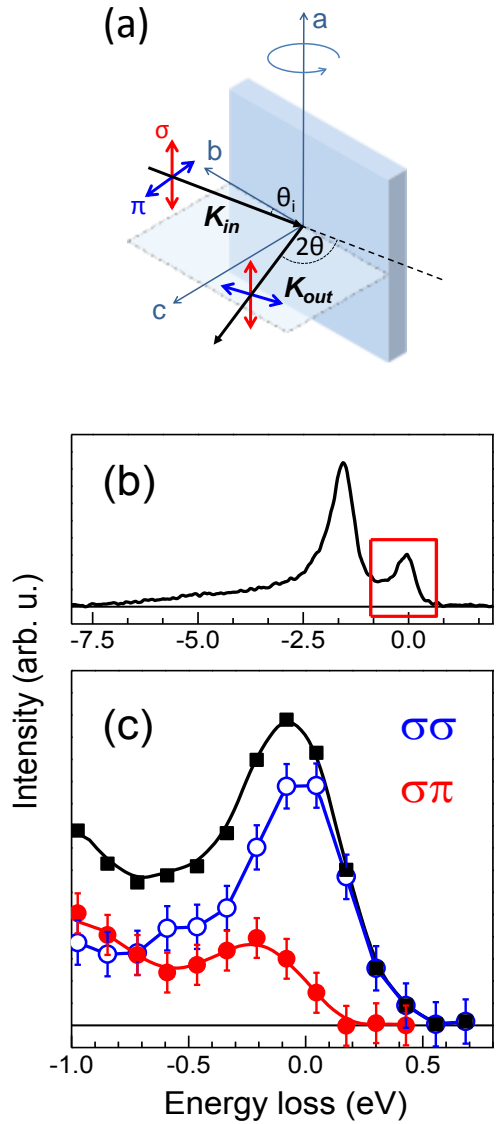

FIGURE 7. Polarimeter results on underdoped $YBa_2Cu_3O_{6.6}$ (hole concentration $p = 0.12$) at 30 deg grazing incidence and σ incident polarization. A schematic cartoon of the experimental geometry is presented in panel (a). The traditional spectrum integrated over the outgoing polarization is given in panel (b). The spectral function within the red box is expanded in panel (c) (black squares) giving the decomposition in the crossed polarization component (red circles) and in non-crossed component (blue open circles).

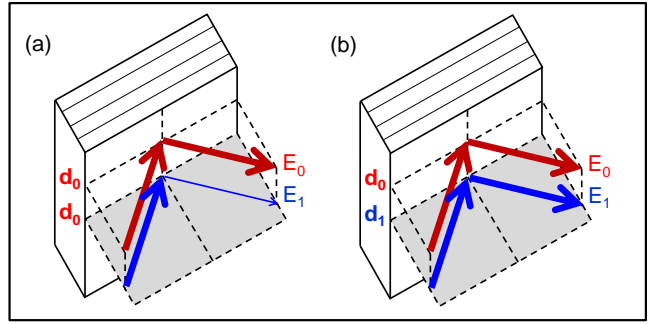

FIGURE 8. Multilayer top view in case of collimated beam (the thickness of the arrows indicates the intensity). Panel (a) (multilayer with uniform spacing): $E_0$ is at Bragg condition of the ML, while $E_1$ (with $E_1 > E_0$) is out of tune and is reflected with lower efficiency. Panel (b) (graded multilayer): the ML in the $E_1$ position is now tuned to $E_1$ thanks to a different period $d$, and the response function is almost flat.